\documentclass{JHEP3}
\newcommand{\beq}{\begin{equation}}
\newcommand{\eeq}{\end{equation}}
\newcommand{\phib}{\ensuremath{\overline{\phi}}}
\newcommand{\etab}{\ensuremath{\overline{\eta}}}
\newcommand{\psib}{\ensuremath{\overline{\psi}}}

\newcommand{\KD}{K\"{a}hler-Dirac }

\usepackage{epsfig,multicol}

\title{Lattice formulation of ${\cal N}=4$ super Yang-Mills theory}

\author{Simon Catterall\\
        Department of Physics, Syracuse University, Syracuse, NY 13244, USA\\
        E-mail: \email{smc@physics.syr.edu}\\
        }

\preprint{SU-4252-808}  

\abstract{We construct a lattice action for ${\cal N}=4$ super Yang-Mills
theory in four dimensions 
which is local, gauge invariant, free of spectrum doubling and
possesses a single exact supersymmetry.
Our construction starts from
the observation that the fermions of the continuum theory can be mapped
into the component fields of a single real anticommuting \KD field.
The original supersymmetry algebra then implies the existence of a 
nilpotent scalar supercharge $Q$ and a corresponding set of
bosonic superpartners. Using this
field content we write down a $Q$-exact action
and show that, with an appropriate change of variables, it reduces
to a well-known {\it twist} of ${\cal N}=4$ super Yang-Mills theory
due to Marcus.  
Using the discretization prescription
developed earlier \cite{2dsuper} we are able to translate this
geometrical action to the lattice. 
}


\begin{document} 
\section{Introduction}
Attempts to formulate lattice supersymmetric theories have a long
history (see \cite{old} and the recent reviews by Feo
and Kaplan \cite{feo_rev,kap_rev}).
Recently there has been renewed interest in this problem stemming from
the realization that in certain classes of theory it may be possible to
preserve at least some of the supersymmetry {\it exactly} at finite
lattice spacing. It is hoped that this residual supersymmetry may
protect the lattice theory from dangerous SUSY-violating radiative corrections.
In the case of
extended supersymmetry two approaches have been followed\footnote{
For examples of ${\cal N}=1$ models with exact SUSY see the recent
work \cite{feowz} and \cite{moto}}; in the first the lattice theory is
constructed by orbifolding a certain supersymmetric matrix model
\cite{kap,others}.
The second approach relies on reformulating the supersymmetric theory in
terms of a new set of variables -- the {\it twisted} fields. In this
procedure a scalar nilpotent supercharge is exposed and it is the algebra
of this charge that one may hope to preserve under discretization \cite{top}.
This approach was initially used to construct lattice formulations of a variety of
low dimensional theories without gauge symmetry \cite{qm,wz2,sigma}. A possible
generalization to lattice gauge theories was given by Sugino \cite{sug}. However,
Sugino's models in four dimensions suffer from the presence of additional states
which do not appear to decouple in the limit of vanishing lattice spacing.

In this paper we introduce a new discretization of the ${\cal N}=4$
twisted Yang-Mills
action which is a generalization of the procedure used earlier to
construct a lattice theory of ${\cal N}=2$ super Yang-Mills theory in
two dimensions \cite{2dsuper}. 
The approach emphasizes the geometrical character of the twisted theory -- the twist
of ${\cal N}=4$ that we consider contains only integer spin fields and the
fermion content is naturally embedded in a (real) \KD field. The connection between
twisting and the \KD fermion mechanism has been emphasized in recent papers
by Kawamoto et al.\cite{noboru}. The idea of using \KD fermions in
order to formulate lattice supersymmetry was first proposed
in \cite{schwim}. In this way a scalar supercharge $Q$ is produced and
the bosonic field content of the model may be embedded in another \KD field. 
Using this field content we write down a $Q$-exact geometrical
action which, after a simple
change of variables, we show is nothing more but a well-known twist of
${\cal N}=4$ super Yang-Mills. We further show explicitly how to recover the
conventional formulation involving spinor fields from the twisted action
showing the complete equivalence of the twisted and untwisted 
theories.\footnote{In this paper we only consider theories
in {\it flat} spacetime and there is thus no distinction
between upper and lower indices for tensors}

Finally, using a discretization prescription
developed earlier, we are able to translate
this geometrical theory to
a hypercubic lattice while preserving gauge invariance and the twisted supersymmetry
and {\it without} inducing any spurious zeroes in the spectrum of the lattice fermion
operator. The price we pay for this is that the lattice theory requires a 
complexification of the degrees of freedom. We conjecture that we can restrict
the path integrals needed to define the Euclidean theory to the
real line while preserving the Ward identities associated to
the $Q$-symmetry.

\section{Twisting and \KD fields}
Consider the field content of ${\cal N}=4$
super Yang-Mills. It can be written in terms of $4$ Majorana spinors $\Psi_\alpha^i$
where the index $\alpha$ labels the spinor degrees of
freedom and the index $i=1\ldots 4$ and 
is associated with an $SO(4)_R$ R-symmetry of the fermionic action. The twisting
procedure consists of constructing a new rotation group which is the diagonal
subgroup of the original Euclidean rotation group $SO(4)$ and this R-symmetry
\cite{noboru,witten,conf,schwim}.
\beq
SO(4)^\prime={\rm diag}\left(SO(4)\times SO(4)_R\right)\eeq
This implies that the two indices $(\alpha ,i)$ should be taken as equivalent
and the fermion field is to be regarded as a matrix $\Psi^i_\alpha\to\Psi_{\beta\alpha}$.
It is then natural to expand this matrix on a basis of products of
$\gamma$-matrices
\beq
\Psi=\eta I+\psi_\mu\gamma_\mu+\frac{1}{2!}\chi_{\mu\nu}\gamma_\mu\gamma_\nu+
\frac{1}{3!}\theta_{\mu\nu\lambda}\gamma_\mu\gamma_\nu\gamma_\lambda+
\frac{1}{4!}\kappa_{\mu\nu\lambda\rho}\gamma_\mu\gamma_\nu\gamma_\lambda\gamma_\rho\eeq
The coefficients $\eta$, $\psi_\mu$ etc are the {\it twisted} fields and
clearly correspond to grassman valued tensors antisymmetric under the exchange of any
two indices. There are exactly $16$ independent fields in this
decomposition and thus it is natural to take all these tensors (or p-forms) as real
to match the $16$ real supercharges of the original theory.

We can then assemble the component fields into a single so-called
\KD field 
$\Psi=\left(\eta,\psi_\mu,\chi_{\mu\nu},\theta_{\mu\nu\lambda},
\kappa_{\mu\nu\lambda\rho}\right)$
It is then straightforward to show that solutions
of the Dirac equation for four degenerate fermions (corresponding
to the four columns of the matrix $\Psi$) can be gotten by
solving the
\KD equation \cite{banks,rabin,betch,joos}
\beq
\left(d-d^\dagger\right)\Psi=0\eeq
where the action of the exterior derivative operator $d$ on a $p$-form $\alpha$
yields a $(p+1)$-form $\beta$ whose components are given by
\beq
\beta_{\mu_1\ldots\mu_{p+1}} =
\partial_{\left[\mu_{p+1}\right.}\alpha_{\left.\mu_1\ldots\mu_p\right]}\eeq
and the square bracket notation for the subscripts indicates an antisymmetrization
with respect to all pairs of indices. The dot product of two such $p$-forms
is defined by
\beq
<\alpha .\beta>=\int dV \frac{1}{p!} 
\alpha_{\mu_1\ldots\mu_p}\beta_{\mu_1\ldots\mu_p}\eeq
With
respect to this dot product we can then define
the adjoint operator $-d^\dagger$ whose action
on a $p$-form $\alpha$ yields a $(p-1)$-form $\beta$ with components
\beq
\beta_{\mu_2\ldots\mu_{p}} =
\partial_{\mu_1}\alpha_{\mu_1\ldots\mu_p}\eeq
These results also hold when the usual derivative is replaced by a gauge
covariant derivative and all fields take values in the adjoint representation
of some $U(N)$ gauge group. It is also straightforward to verify
that the \KD equation can be obtained from a \KD action of the form
\beq S_{\rm KD}=<\Psi^\dagger .\left(d-d^\dagger\right) \Psi>\eeq
or equivalently in matrix language
\beq S_{\rm KD}=\frac{1}{2}{\rm Tr}\Psi^\dagger \gamma .D \Psi\eeq
 
This representation of fermions in terms of $p$-forms is very natural
from the lattice perspective as the latter may be associated with
lattice $p$-cochains --
functions defined on $p$-dimensional simplices in the lattice.
Lattice analogs of the exterior derivative and its adjoint exist
and allow us to discretize continuum actions formulated in
geometric terms in a well-defined way. One of the most important
consequences of such discretizations is that they prohibit
spectrum doubling -- the appearance of spurious zeroes of the
fermion operator associated with lattice modes which do not
appear in the continuum theory \cite{rabin,betch,joos}.

Notice that this twisting procedure will yield a scalar supercharge $Q$ which implies
that the twisted theory will also contain a set of corresponding commuting p-form fields
$\Phi=\left(\phib,A_\mu,B_{\mu\nu},W_{\mu\nu\lambda},C_{\mu\nu\lambda\rho}\right)$.
The fields $B_{\mu\nu}$ and $C_{\mu\nu\lambda\rho}$ will turn out to be
multiplier fields which are integrated out of the final theory, leaving the
bosonic fields to be represented by the gauge field $A_\mu$, the scalar
$\phib$, another scalar gauge parameter $\phi$ and the four independent degrees of freedom carried
by $W_{\mu\nu\lambda}$. It is clear that in the twisting process the original
6 scalar fields have decomposed into a $\bf 4+1+1$ of $SO(4)^\prime$. 
The appropriate action of $Q$ on these fields is a simple generalization of the
two dimensional ${\cal N}=2$ case\footnote{Notice the minus sign which appears
in the $Q$-variation of $\psi_\mu$ which was missing in our earlier
paper \cite{2dsuper} -- many thanks to Mithat Unsal for pointing out this
error}
\begin{eqnarray}
Q\phib&=&\eta\;\;Q\eta=[\phi,\phib]\nonumber\\
QA_\mu&=&\psi_\mu\;\;Q\psi_\mu=-D_\mu\phi\nonumber\\
QB_{\mu\nu}&=&[\phi,\chi_{\mu\nu}]\;\;Q\chi_{\mu\nu}=B_{\mu\nu}\nonumber\\
QW_{\mu\nu\lambda}&=&\theta_{\mu\nu\lambda}\;\;Q\theta_{\mu\nu\lambda}=
[\phi,W_{\mu\nu\lambda}]\nonumber\\
QC_{\mu\nu\lambda\rho}&=&[\phi,\kappa_{\mu\nu\lambda\rho}]\;\;Q\kappa_{\mu\nu\lambda\rho}=
C_{\mu\nu\lambda\rho}\nonumber\\
Q\phi&=&0
\label{Q}
\end{eqnarray}
Notice that, as expected from the twisted
superalgebra \cite{conf}, this charge is nilpotent up to a gauge transformation
parametrized by the additional scalar field $\phi$ -- $Q^2=\delta_G^\phi$. 
Since the entire field content of this twisted model is given in terms of
p-form fields it will be natural that only the exterior derivative and its
adjoint may appear in the action of the twisted theory. This will guide us in
the construction of the appropriate action. 

\section{Continuum Action}
\subsection{Geometric Formulation}
We will hypothesize that the action can be written in a $Q$-exact form as for
the two dimensional ${\cal N}=2$ theory. Thus $S=\beta Q\Lambda$ where $\beta$ is
a coupling and $\Lambda(\Psi,\Phi)$ will be termed a gauge fermion in agreement
with the usual BRST terminology. All fields should be regarded as expanded on
a basis of antihermitian generators of $U(N)$.
We choose $\Lambda$ to be of the form
\begin{eqnarray}
\Lambda&=&\int d^4x {\rm Tr}\left[
\chi_{\mu\nu}\left(F_{\mu\nu}+\frac{1}{2}B_{\mu\nu}-
\frac{1}{2}[W_{\mu\lambda\rho},W_{\nu\lambda\rho}]+
D_\lambda W_{\lambda\mu\nu}\right)\right.\nonumber\\
&+&\left.\psi_\mu D_\mu\phib+\frac{1}{4}\eta[\phi,\phib]+
\frac{1}{3!}\theta_{\mu\nu\lambda}[W_{\mu\nu\lambda},\phib]\right.\nonumber\\
&+&\left.\frac{1}{4!}\kappa_{\mu\nu\lambda\rho}\left(
\sqrt{2}D_{\left[\mu\right.}W_{\left.\nu\lambda\rho\right]}+
\frac{1}{2}C_{\mu\nu\lambda\rho}\right)
\right]
\label{qact}
\end{eqnarray}
Several of these terms are in common with the gauge fermion of ${\cal N}=2$ super
Yang-Mills theory in two dimensions. The new ones involve the $3$ and $4$-form
fields. Of these the terms involving derivatives must be present to generate the
correct \KD action for the twisted fermions (and will simultaneously
generate the appropriate kinetic terms for
the W-field).
The commutator term involving $W_{\mu\nu\lambda}$ coupled to
$\chi_{\mu\nu}$ will generate a quartic potential for the W-field analogous to
that generated for the scalars $\phi$ and $\phib$. This will allow contact
to be made eventually with the supersymmetric theory
where one expects the scalars and W-field to play similar roles. 
In the same way the
commutator term involving $\phib$ and $W$ will also generate the
necessary mixed quartic
couplings between the scalars and the W-field.
Carrying out the $Q$-variation leads to the following action
\beq S=\beta\left(S_B+S_F+S_Y\right)\eeq
where the piece of the action $S_B$ involving the bosonic fields takes the form
\begin{eqnarray}
S_B&=&\int d^4x {\rm Tr}\left[
B_{\mu\nu}\left(F_{\mu\nu}-
\frac{1}{2}[W_{\mu\lambda\rho},W_{\nu\lambda\rho}]+
D_\lambda W_{\lambda\mu\nu}+
\frac{1}{2}B_{\mu\nu}\right)\right.\nonumber\\
&-&\left.D_\mu\phi D_\mu\phib+\frac{1}{4}[\phi,\phib]^2-
\frac{1}{3!}[\phi,W_{\mu\nu\lambda}][\phib,W_{\mu\nu\lambda}]\right.\nonumber\\
&+&\left.\frac{1}{4!}C_{\mu\nu\lambda\rho}\left(
\sqrt{2}D_{\left[\mu\right.}W_{\left.\nu\lambda\rho\right]}+
\frac{1}{2}C_{\mu\nu\lambda\rho}\right)
\right]
\end{eqnarray}
and the fermion kinetic terms are given by $S_F$ with
\begin{eqnarray}
S_F&=&\int d^4 x {\rm Tr}\left[
-\chi_{\mu\nu}D_{\left[\mu\right.}\psi_{\left.\nu\right]}
-\chi_{\mu\nu}D_\lambda\theta_{\lambda\mu\nu}
-\eta D_\mu\psi_\mu-
\frac{\sqrt{2}}{4!}\kappa_{\mu\nu\lambda\rho}D_{\left[\mu\right.}
\theta_{\left.\nu\lambda\rho\right]}
\right]
\end{eqnarray}
and $S_Y$ contains the Yukawa couplings
\begin{eqnarray}
S_Y&=&\int d^4x {\rm Tr}\left[
-\frac{1}{4}\eta[\phi,\eta]
-\frac{1}{2}\frac{1}{4!}
\kappa_{\mu\nu\lambda\rho}[\phi,\kappa_{\mu\nu\lambda\rho}]-
\frac{1}{2}\chi_{\mu\nu}[\phi,\chi_{\mu\nu}]\right.\nonumber\\
&+&\left.\psi_\mu[\phib,\psi_\mu]+
\frac{1}{3!}\theta_{\mu\nu\lambda}[\phib,\theta_{\mu\nu\lambda}]
\right.\nonumber\\
&+&\left.\frac{1}{3!}\eta[\theta_{\mu\nu\lambda},W_{\mu\nu\lambda}]-
\frac{\sqrt{2}}{4!}\kappa_{\mu\nu\lambda\rho}
[\psi_{\left[\mu\right.},W_{\left.\nu\lambda\rho\right]}]\right.\nonumber\\
&+&\left.\chi_{\mu\nu}[\theta_{\mu\lambda\rho},W_{\nu\lambda\rho}]-
\chi_{\mu\nu}[\psi_\lambda,W_{\lambda\mu\nu}]
\right]
\end{eqnarray}
Integrating over the multiplier fields $B_{\mu\nu}$ and $C_{\mu\nu\lambda\rho}$
and subsequently 
utilizing the Bianchi identity leads to a new bosonic action
of the form
\begin{eqnarray}
S_B&=&\int d^4 x {\rm Tr}\left[
-\frac{1}{2}\left(
\left(F_{\mu\nu}-
\frac{1}{2}[W_{\mu\lambda\rho},W_{\nu\lambda\rho}]\right)^2
+\left(D_\lambda W_{\lambda\mu\nu}\right)^2+
\frac{2}{4!}\left(D_{\left[\mu\right.}
W_{\left.\nu\lambda\rho\right]}\right)^2\right)
\right.\nonumber\\
&-&\left.D_\mu\phi D_\mu\phib+\frac{1}{4}[\phi,\phib]^2-
\frac{1}{3!}[\phi,W_{\mu\nu\lambda}][\phib,W_{\mu\nu\lambda}]
\right]
\end{eqnarray}

\subsection{Relation to the Marcus twist of ${\cal N}=4$ SYM}
At this point it is useful to trade the $W$, $\theta$, and $\kappa$ fields
for new variables which will allow contact to be made between this
theory and one of the conventional twists of ${\cal N}=4$ super Yang-Mills.
We write
\begin{eqnarray}
W_{\mu\nu\lambda}&=&\epsilon_{\mu\nu\lambda\rho}V_\rho\nonumber\\
\theta_{\mu\nu\lambda}&=&\epsilon_{\mu\nu\lambda\rho}\psib_\rho\nonumber\\
\kappa_{\mu\nu\lambda\rho}&=&\epsilon_{\mu\nu\lambda\rho}\etab
\end{eqnarray}
In terms of these variables the bosonic action reads
\begin{eqnarray}
S_B&=&\int d^4 x {\rm Tr}\left[
-\frac{1}{2}\left(
\left(F_{\mu\nu}-
[V_{\mu},V_{\nu}]\right)^2
+\left(D_{\left[\mu\right.}V_{\left.\nu\right]}\right)^2+
2\left(D_\mu V_\mu\right)^2
\right)
\right.\nonumber\\
&-&\left.D_\mu\phi D_\mu\phib+\frac{1}{4}[\phi,\phib]^2-
[\phi,V_{\mu}][\phib,V_{\mu}]
\right]
\end{eqnarray}
A further integration by parts yields a cancellation between
the terms linear in $F_{\mu\nu}$ and the final bosonic action becomes
\begin{eqnarray}
S_B&=&\int d^4x {\rm Tr}\left[
-\frac{1}{2}F^2_{\mu\nu}-\frac{1}{2}[V_\mu,V_\nu]^2
-\left(D_\mu V_\nu\right)^2\right.\nonumber\\
&-&\left.D_\mu\phi D_\mu\phib+\frac{1}{4}[\phi,\phib]^2-
[\phi,V_{\mu}][\phib,V_{\mu}]
\right]
\end{eqnarray}
Making the additional rescalings $\chi\to 2\chi$ and $\etab\to \frac{1}{\sqrt{2}}\etab$ we find the fermion kinetic term takes the form
\begin{equation}
S_F=\int d^4 x {\rm Tr}\left[
-2\chi_{\mu\nu}D_{\left[\mu\right.}\psi_{\left.\nu\right]}
-2\chi^*_{\mu\nu}D_{\left[\mu\right.}\psib_{\left.\nu\right]}
-2\frac{\eta}{2} D_\mu\psi_\mu
-2\frac{\etab}{2} D_\mu\psib_\mu
\right]
\end{equation}
where 
$\chi^*_{\mu\nu}=\frac{1}{2}\epsilon_{\mu\nu\lambda\rho}\chi_{\lambda\rho}$
is the dual field.
In these variables the Yukawa's take on the more symmetrical
form
\begin{eqnarray}
S_Y&=&\int d^4x {\rm Tr}\left[
-\frac{\eta}{2}[\phi,\frac{\eta}{2}]
-\frac{\etab}{2}[\phi,\frac{\etab}{2}]
-2\chi_{\mu\nu}[\phi,\chi_{\mu\nu}]\right.\nonumber\\
&+&\left.\psi_\mu[\phib,\psi_\mu]+
\psib_\mu[\phib,\psib_\mu]
\right.\nonumber\\
&+&\left.2\frac{\eta}{2}[\psib_{\mu},V_\mu]-2\frac{\etab}{2}[\psi_\mu,V_\mu]
\right.\nonumber\\
&+&4\left.\chi_{\mu\nu}[\psib_{\mu},V_{\nu}]-
4\chi^*_{\mu\nu}[\psi_\mu,V_{\nu}]
\right]
\end{eqnarray}
The new action can be recognized as the twist of ${\cal N}=4$ super
Yang-Mills due to Marcus \cite{marcus}. This is made explicit by the further
change of variables
\begin{eqnarray}
\eta_M&=&\frac{1}{2}\left(\eta-i\etab\right)\nonumber\\
\psi_M&=&\frac{1}{2}\left(\psi-i\psib\right)\nonumber\\
\chi_M&=&2\left(\chi-i\chi^*\right)\nonumber\\
B_M&=&\frac{1}{2}\phi\nonumber\\
C_M&=&\phib
\end{eqnarray}
Notice that this twist of ${\cal N}=4$ super Yang-Mills was also
analyzed in \cite{lambastida} although in that paper two scalar supercharges
were constructed which transformed into other under a duality operation.
We will see later that such duality operations are incompatible with
our latticization prescription and only the single supercharge 
corresponding to the transformations in equation
\ref{Q} can be adapted to the lattice. It is interesting that this
twist of ${\cal N}=4$ can also be interpreted as a deformation of
four dimensional super BF-theory \cite{blau}.

\subsection{Connection to conventional formulation of ${\cal N}=4$ SYM}
Finally we will show how this twisted model may be reinterpreted in
terms of the usual formulation of ${\cal N}=4$ super Yang-Mills theory.
First, concentrate on the bosonic action and introduce the new
fields ($\phi=\phi_1+i\phi_2$)
\begin{eqnarray}
X^\mu&=&V_\mu\;\;\mu=0\ldots 3\nonumber\\
X^4&=&\phi_1\nonumber\\
X^5&=&\phi_2
\end{eqnarray}
Then the bosonic action may be trivially rewritten
as
\beq
S_B=-\frac{1}{2}F^2_{\mu\nu}-\left(D_\mu X^i\right)^2-
\frac{1}{2}\sum_{ij}[X_i,X_j]^2
\eeq
Notice this is real, positive semidefinite on account of the
antihermitian basis for the fields. It is also precisely the bosonic
sector of the ${\cal N}=4$ super Yang-Mills action with $X^i$
the usual $6$ real scalars of that theory. 

Next let us turn our attention to the fermion kinetic term.
From our previous discussion 
it should be clear that the twisted fermion kinetic term is nothing
more than the component expansion of a \KD action:
\beq S_F=\frac{1}{2}{\rm Tr}\Psi^\dagger \gamma .D \Psi \eeq
where $\Psi$ is the \KD field defined earlier with $\eta\to\eta/2$, 
$\theta_{\mu\nu\lambda}\to\epsilon_{\mu\nu\lambda\rho}\psib_\rho$
and $\kappa_{\mu\nu\lambda\rho}\to\epsilon_{\mu\nu\lambda\rho}\etab/2$.
Naively such an action appears to describe a theory with {\it four}
Dirac spinor fields - rather than the two one would have expected
for ${\cal N}=4$ super Yang-Mills. However it is evident that $\Psi$
obeys a reality condition if its associated \KD field is real. This
reduces the action to that of two degenerate
Dirac fermions. To see this in detail let us 
adopt a Euclidean chiral basis for the $\gamma$-matrices
\beq
\begin{array}{cc}
\gamma_0=\left(\begin{array}{cc}
0&I\\
I&0\end{array}\right)
&
\gamma_i=\left(\begin{array}{cc}
0&-i\sigma_i\\
i\sigma_i&0\end{array}\right)
\end{array}
\eeq
It is straightforward to see that the $\gamma$ matrices (and any
products of them) obey a reality condition
\beq
\gamma_\mu^*=C\gamma_\mu C^{-1}\eeq
where the matrix $C$ is given by
\beq C=\left(\begin{array}{cc}
\sigma_2&0\\
0&\sigma_2\end{array}\right)\eeq
With real p-form coefficients this implies
a reality condition on $\Psi$ itself
\beq \Psi^*=C\Psi C^{-1}\eeq
This in turn implies that $\Psi^\dagger=C\Psi^T C^{-1}$ which makes it
clear that the result of integrating over the
\KD field $\Psi$ should be the Pfaffian of the \KD operator
(let us neglect the Yukawa couplings for the moment).
This, in turn, will correspond to the product of {\it two} 
Dirac determinants.

To see this in more detail one can use the
reality condition to show that successive columns $\Psi^{(n)}$ of $\Psi$ 
are not independent but are charge conjugates of each other
\beq \Psi^{(2)}=C\left(\Psi^{(1)}\right)^*\;\;
\Psi^{(4)}=C\left(\Psi^{(3)}\right)^*\eeq
These conditions allow us to rewrite
the twisted fermion kinetic term in the conventional form
\beq \frac{1}{2}\sum_{\alpha=1,2}\lambda^{\dagger}_\alpha \gamma .D \lambda_\alpha\eeq
where the spinors are read off as the first and third columns of
the $\Psi$ matrix in this chiral basis:
\beq
\begin{array}{cc}
\lambda_1=\left(\begin{array}{c}
\frac{\eta}{2}-\frac{\etab}{2}+2i\chi^+_{03}\\
-2\chi^+_{02}+2i\chi^+_{01}\\
\left(\psi_0+\psib_0\right)+i\left(\psi_3+\psib_3\right)\\
-\left(\psi_2+\psib_2\right)+i\left(\psi_1+\psib_1\right)
\end{array}\right)
&
\lambda_2=\left(\begin{array}{c}
\left(\psi_0-\psib_0\right)+i\left(-\psi_3+\psib_3\right)\\
\left(\psi_2-\psib_2\right)+i\left(\psib_1-\psi_1\right)\\
\frac{\eta}{2}+\frac{\etab}{2}-2i\chi^-_{03}\\
2\chi^-_{02}-2i\chi^-_{01}
\end{array}\right)\\
\end{array}
\eeq
Here, $\chi^{\pm}=\frac{1}{2}\left(\chi\pm\chi^*\right)$ are the usual
self-dual and antiself-dual parts of the field. 
The Yukawa's can also be put in the general form
\beq \frac{1}{2}\sum_{\alpha=1,2}\lambda^\dagger_\alpha
C^\alpha_i\Gamma^i[X^i,\lambda_\alpha]\eeq
where $\Gamma^i=\left\{I,\gamma_5,\gamma_\mu\gamma_5,\mu=0\ldots 4\right\}$
and the coefficients $C^\alpha_i$ are just numbers. This is just
the structure expected of ${\cal N}=4$ super Yang-Mills.
Notice that both Dirac operators $M^\alpha$ including the Yukawa
interactions possess the symmetry
\beq \left(D^\alpha\right)^*=C D^\alpha C^{-1}\eeq
which shows that their eigenvalves come in complex conjugate
pairs and hence the associated determinants are positive definite.

Up to this point we have shown that the original $Q$-exact action
written in terms of the component tensors of \KD fields \ref{qact}
may be rewritten in terms of a conventional twisted action which may
in turn be used to reconstruct {\it exactly} the usual formulations
of ${\cal N}=4$ super Yang-Mills theory. All this has been in the
continuum. The recasting of the theory in terms of these
geometrical fields is crucially important when devising a 
transcription to the lattice which preserves as much as possible of
the continuum symmetry. We turn to this now.

\section{Lattice Action}
\subsection{Geometric Formulation}
The lattice action is obtained by discretization of the 
$Q$-exact action given in eqn.~\ref{qact}. The prescription we
employ was introduced in our earlier paper on ${\cal N}=2$ super
Yang-Mills in two dimensions and draws on the
work in \cite{adjoint,geo}. It is summarized below
\begin{itemize}
\item A continuum p-form field $f_{\mu_1\ldots \mu_p}(x)$ will be mapped
to a corresponding lattice field associated with
the $p$-dimensional hypercube at lattice site $x$ 
spanned by the (positively directed) unit vectors $\{\mu_1\ldots\mu_p\}$. 
\item Such a lattice
field is taken to transform under gauge transformations in the following way
\beq
f_{\mu_1\ldots\mu_p}(x)\to G(x)f_{\mu_1\ldots \mu_p}(x)G^{-1}(x+e_{\mu_1\ldots\mu_p})\eeq
where the vector $e_{\mu_1\ldots\mu_p}=\sum_{j=1}^p\mu_j$. 
\item To construct gauge invariant quantities we will need to introduce both
$f_{\mu_1\ldots\mu_p}$ 
and its hermitian conjugate $f^\dagger_{\mu_1\ldots\mu_p}(x)$. The
latter transforms as
\beq
f^\dagger_{\mu_1\ldots\mu_p}(x)\to G(x+e_{\mu_1\ldots\mu_p})
f_{\mu_1\ldots \mu_p}(x)G^{-1}(x)\eeq
Since for all fields bar the gauge field we will assume
$f_{\mu_1\ldots\mu_p}=\sum_a f^a_{\mu_1\ldots\mu_p}T^a$ where
$T^a$ are antihermitian generators of $U(N)$, this
necessitates taking the fields $f^a_{\mu_1\ldots\mu_p}$ to be
{\it complex}.
\item For a continuum gauge field we introduce lattice link 
fields $U_\mu(x)=e^{A_\mu(x)}=e^{A^a_\mu(x)T^a}$ with $A^a_\mu(x)$
complex together with 
its conjugate
$U^\dagger_\mu=e^{A^\dagger_\mu(x)}$. 
\item A covariant forward difference operator can be defined which acts on a field
$f_{\mu_1\ldots\mu_p}(x)$ as follows
\beq
D^+_\mu f_{\mu_1\ldots\mu_p}(x)=
U_\mu(x)f_{\mu_1\ldots\mu_p}(x+\mu)-
f_{\mu_1\ldots\mu_p}(x)U_\mu(x+e_{\mu_1\ldots\mu_p})\eeq
This operator acts like a lattice exterior derivative with
respect to gauge transformations in mapping a $p$-form lattice field
to a $(p+1)$-form lattice field.
\item Similarly we can define an adjoint operator $D^-_\mu$ 
whose action on some field
$f_{\mu_1\ldots\mu_p}$ is given by
\beq
D^-_\mu f_{\mu_1\ldots\mu_p}(x)=
f_{\mu_1\ldots\mu_p}(x)U^\dagger_\mu(x+e_{\mu_1\ldots\mu_p}-\mu)-
U^\dagger_\mu(x-\mu)f_{\mu_1\ldots\mu_p}(x-\mu)\eeq
It thus acts like the adjoint of the exterior derivative.
\item Following \cite{rabin} all instances of $\partial_\mu$ in the
continuum action will be replaced by $D^+_\mu$ if the derivative
acts like $d$ (curl-like operation) and $D^-_\mu$ if the derivative acts 
like $d^\dagger$ (divergence-like operation).
One can show using results from homology theory \cite{rabin} that
this guarantees that the lattice theory will exhibit no spectrum
doubling.
\end{itemize}
Notice that each $p$-hypercube (for $p>0$) possesses two
orientations and this gives one natural explanation of the
doubling of degrees of freedom exhibited by the lattice
theory\footnote{Another possible interpretation relates to
the existence of both lattice and dual lattice}. The need for
this doubling can be easily seen at an
operational level -- with the lattice gauge transformation rules 
we have given the lattice operator
\beq
\int d^4x A_{\mu_1\ldots\mu_d}B_{\mu_1\ldots\mu_d}\eeq
is {\it not} gauge invariant (and if used in the gauge fermion
will also lead to a violation of $Q$-invariance). However, there is a 
natural way to construct a gauge invariant lattice operator
\beq
\int d^4x A^\dagger_{\mu_1\ldots\mu_d}B_{\mu_1\ldots\mu_d}\eeq
However, the differing gauge transformations of 
$A^\dagger_{\mu_1\ldots\mu_d}$ and
$A_{\mu_1\ldots\mu_d}$ then require that the component fields
$A^a_{\mu_1\ldots\mu_d}$ be taken as complex. Clearly this
prescription must be used consistently for all fields and
necessitates treating the gauge links as {\it non-unitary} matrices.
This rule naturally leads to a gauge
action of the form $F^\dagger_{\mu\nu}F_{\mu\nu}$ which, we
will see later, reduces
to the usual Wilson action in the unitary limit -- a feature
which would not have occurred with the
lattice operator $F_{\mu\nu}F_{\mu\nu}$.

Notice that our gauge transformation
rules allow for a complex gauge transformation parameter $\phi(x)$.
This is natural in such a complexified theory. Compatibility with
the dagger operation requires that $\phi^\dagger=-\phi$ which should
be true for {\it all} scalar fields.
While
the symmetries are most easily implemented in the complexified theory
we will argue that the final path integral can be restricted to
the real line without violating either gauge invariance or
the twisted supersymmetry. 

Using these ingredients we can easily transfer the continuum gauge fermion
in eqn.~\ref{qact} to the lattice obtaining
\begin{eqnarray}
\Lambda&=&\frac{1}{2}\sum_x {\rm Tr}\left[
-\chi^\dagger_{\mu\nu}\left(F_{\mu\nu}+\frac{1}{2}B_{\mu\nu}-
\frac{1}{2}[W_{\mu\lambda\rho},W_{\nu\lambda\rho}]^\prime+
D^-_\lambda W_{\lambda\mu\nu}\right)\right.\nonumber\\
&-&\left.\psi^\dagger_\mu D^+_\mu\phib-\frac{1}{4}\eta^\dagger[\phi,\phib]-
\frac{1}{3!}\theta^\dagger_{\mu\nu\lambda}[W_{\mu\nu\lambda},\phib]\right.\nonumber\\
&-&\left.\frac{1}{4!}\kappa^\dagger_{\mu\nu\lambda\rho}\left(
\sqrt{2}D^+_{\left[\mu\right.}W_{\left.\nu\lambda\rho\right]}+
\frac{1}{2}C_{\mu\nu\lambda\rho}\right)+{\rm h.c}
\right]
\label{qact_lat}
\end{eqnarray}
Notice that it is necessary to add the hermitian conjugates of these terms
to the gauge fermion
to obtain a real lattice action. 
The extra minus signs just reflect the antihermitian
nature of our basis matrices $T^a$.
Unlike the continuum theory we are
no longer at liberty to transform $3$-form and $4$-form fields to
their duals since this would break lattice gauge invariance.
The lattice field strength $F_{\mu\nu}$
is obtained from the simple relation
\beq
F_{\mu\nu}(x)=D^+_\mu U_\nu(x)\eeq
Notice that it automatically antisymmetric in its indices and reduces
to the usual Yang-Mills field strength in the naive continuum
limit. The prime on the commutator 
of two $W$-fields indicates that this term must be modified in the
lattice theory in order to maintain gauge invariance. The following
definition of the commutator yields a term which transforms as a lattice 2-form
\beq
\left\{W_{\mu\lambda\rho}(x)W_{\nu\lambda\rho}(x+\mu+\lambda+\rho)-
W_{\nu\lambda\rho}(x)W_{\mu\lambda\rho}(x+\nu+\lambda+\rho)\right\}
P(x+\mu+\nu;\lambda,\rho)\eeq
where the ordered link path $P(x+\mu+\nu;\lambda,\rho)$ 
is necessary to ensure that the gauge
transformation properties of the commutator do not depend on the
indices $\lambda$ and $\rho$.
\beq
P(x;\lambda,\rho)=U^\dagger_\lambda(x+\lambda+2\rho)
U^\dagger_\lambda(x+2\rho)U^\dagger_\rho(x+\rho)U^\dagger_\rho(x)\eeq
To enforce lattice rotational symmetry one can replace $P(x;\lambda,\rho)$ by
a symmetrized average over all lattice paths leading from 
$(x+2\lambda+2\rho)$ to $x$. Notice that this
lattice commutator term reduces to the continuum commutator in the
naive continuum limit in which the gauge links are set to unity.
Actually there is one further wrinkle to comment on. The commutator
term involving $W$ and $\phib$ must also be modified from its naive
continuum form to maintain gauge invariance. Since this same procedure
must be used in the lattice $Q$-transformations we list the general
rule for the commutator of a scalar field with an arbitrary lattice
$p$-form
\beq
[\phi,f_{\mu_1\ldots\mu_p}]=\phi(x)f_{\mu_1\ldots\mu_p}(x)-
f_{\mu_1\ldots\mu_p}(x)\phi(x+e_{\mu_1\ldots\mu_p})\eeq
The $Q$-transformations listed in eqn.~\ref{Q} 
can now be taken over
almost trivially to the lattice. They take the form
\begin{eqnarray}
Q\phib&=&\eta\;\;Q\eta=[\phi,\phib]\nonumber\\
QU_\mu&=&\psi_\mu\;\;Q\psi_\mu=-D^+_\mu\phi\nonumber\\
QB_{\mu\nu}&=&[\phi,\chi_{\mu\nu}]\;\;Q\chi_{\mu\nu}=B_{\mu\nu}\nonumber\\
QW_{\mu\nu\lambda}&=&\theta_{\mu\nu\lambda}\;\;Q\theta_{\mu\nu\lambda}=
[\phi,W_{\mu\nu\lambda}]\nonumber\\
QC_{\mu\nu\lambda\rho}&=&[\phi,\kappa_{\mu\nu\lambda\rho}]\;\;Q\kappa_{\mu\nu\lambda\rho}=
C_{\mu\nu\lambda\rho}\nonumber\\
Q\phi&=&0
\label{Qlat}
\end{eqnarray}
where gauge invariance dictates the
use of a forward difference in the variation of $\psi_\mu$
and the commutators are to be point split in the way described above
so that they can be interpreted as infinitessimal gauge
transformations on lattice $p$-form fields. For example,
\beq[\phi,\chi_{\mu\nu}]\to
\phi(x)\chi_{\mu\nu}(x)-\chi_{\mu\nu}(x)\phi(x+\mu+\nu)\eeq
The $Q$-transformations of the daggered fields can be found by taking the
hermitian conjugate of these transformations together with the requirement
that scalar fields transform into (minus) themselves under the
dagger operation.

Carrying out the $Q$-variation and subsequently integrating out the multiplier
fields as for the continuum case leads to the following components of the
lattice action
\begin{eqnarray}
S_B&=&\frac{1}{2}\sum_x {\rm Tr}\left[
\left(F_{\mu\nu}-
\frac{1}{2}[W_{\mu\lambda\rho},W_{\nu\lambda\rho}]^\prime+
D^-_\lambda W_{\lambda\mu\nu}\right)
\left(F_{\mu\nu}-
\frac{1}{2}[W_{\mu\lambda\rho},W_{\nu\lambda\rho}]^\prime+
D^-_\lambda W_{\lambda\mu\nu}\right)^\dagger\right.\nonumber\\
&+&\left.\frac{2}{4!}\left(D^+_{\left[\mu\right.}W_{\left.\nu\lambda\rho\right]}\right)
\left(D^+_{\left[\mu\right.}W_{\left.\nu\lambda\rho\right]}\right)^\dagger+
\left(D^+_\mu\phi\right)^\dagger\left(D^+_\mu\phib\right)+
\frac{1}{4}[\phi,\phib]^2
\right.\nonumber\\
&-&\left.\frac{1}{3!}[\phi,W^\dagger_{\mu\nu\lambda}][W_{\mu\nu\lambda},\phib]
+{\rm h.c}\right]
\end{eqnarray}
Notice that the terms listed above are positive semidefinite 
along the contour $\phib =-\phi^\dagger$
corresponding to choosing $\phib^a(x)=\left(\phi^a(x)\right)^*$. Notice also
that the lattice Yang-Mills action $F^\dagger_{\mu\nu}F_{\mu\nu}$ takes the
form 
\beq
{\rm Tr}\sum_x\sum_{\mu<\nu} \left(2I-U^P_{\mu\nu}-(U^P_{\mu\nu})^\dagger\right)+
{\rm Tr}\sum_x\sum_{\mu<\nu} \left(M_{\mu\nu}+M_{\nu\mu}-2I\right)
\eeq
where \beq
U^P_{\mu\nu}={\rm Tr}\left(U_\mu(x)U_\nu(x+\mu)U^\dagger_\mu(x+\nu)U^\dagger_\nu(x)\right)\eeq
is the
usual Wilson plaquette operator
and 
\beq M_{\mu\nu}(x)=U_\mu(x)U^\dagger_\mu(x)U_\nu(x+\mu)U^\dagger_\nu(x+\mu)\eeq
Notice that the latter term vanishes if we restrict the gauge links to be
unitary.
Turning now to the fermion kinetic term we find
\begin{eqnarray}
S_F&=&\sum_x {\rm Tr}\left[
\frac{1}{2!}\chi^\dagger_{\mu\nu}D^+_{\left[\mu\right.}\psi_{\left.\nu\right]}+
\psi^\dagger_{\nu} D^-_{\mu} \chi_{\mu\nu}+
\frac{1}{2!}\chi^\dagger_{\mu\nu}D^-_{\lambda}\theta_{\lambda\mu\nu}+
\frac{1}{3!}\theta^\dagger_{\mu\nu\lambda}D^+_{\left[\mu\right.}
\chi_{\left.\nu\lambda\right]}\right.\nonumber\\
&+&\left.\psi^\dagger_\mu D^+_\mu \frac{\eta}{2}+\frac{\eta^\dagger}{2}D^-_\mu \psi_\mu+
\frac{1}{4!}\frac{\kappa_{\mu\nu\lambda\rho}}{2}
D^+_{\left[\mu\right.}\theta_{\left.\nu\lambda\rho\right]}+
\frac{1}{3!}\theta^\dagger_{\mu\nu\lambda}
D^-_\mu\frac{\kappa_{\mu\nu\lambda\rho}}{2}
\right]\end{eqnarray}
In this expression we have rescaled
$\kappa_{\mu\nu\lambda\rho}\to \frac{1}{\sqrt{2}}\kappa_{\mu\nu\lambda\rho}$
as in the continuum. 
The Yukawa couplings follow in a similar manner
\begin{eqnarray}
S_Y&=&\frac{1}{2}\sum_x {\rm Tr}\left[
\frac{\eta^\dagger}{2}[\phi,\frac{\eta}{2}]
+\frac{1}{4!}\frac{\kappa^\dagger_{\mu\nu\lambda\rho}}{2}[\phi,\frac{\kappa_{\mu\nu\lambda\rho}}{2}]
+\frac{1}{2}\chi^\dagger_{\mu\nu}[\phi,\chi_{\mu\nu}]\right.\nonumber\\
&-&\left.\psi^\dagger_\mu[\phib,\psi_\mu]-
\frac{1}{3!}\theta^\dagger_{\mu\nu\lambda}[\phib,\theta_{\mu\nu\lambda}]
\right.\nonumber\\
&+&\left.2\frac{1}{3!}\theta^\dagger_{\mu\nu\lambda}
[W_{\mu\nu\lambda},\frac{\eta}{2}]+
2\frac{1}{4!}\frac{\kappa^\dagger_{\mu\nu\lambda\rho}}{2}
[\psi_{\left[\mu\right.},W_{\left.\nu\lambda\rho\right]}]
\right.\nonumber\\
&-&\left.\frac{1}{2}\chi^\dagger_{\mu\nu}
[\theta_{\lambda\rho\left[\mu\right.},W_{\left.\nu\right]\lambda\rho}]^\prime
+\chi^\dagger_{\mu\nu}[\psi^\dagger_\lambda,W_{\lambda\mu\nu}]
+{\rm h.c}\right]
\end{eqnarray}
The commutator term involving $\psi_\mu$ and $W_{\nu\lambda\rho}$ is easily
found to take the gauge covariant and point split form
\beq
\left(\psi_{\left[\mu\right.}(x)W_{\left.\nu\lambda\rho\right]}(x+\mu)-
W_{\left[\nu\lambda\rho\right.}(x)\psi_{\left.\mu\right]}(x+\nu+\lambda+\rho)\right)\eeq
Similarly that involving $\psi^\dagger_\lambda$ and $W_{\lambda\mu\nu}$ is found to be
\beq
\left(W_{\lambda\mu\nu}(x)\psi^\dagger_\lambda(x+\mu+\nu)-
\psi^\dagger_\lambda(x-\lambda)W_{\lambda\mu\nu}(x-\lambda)\right)\eeq

There is one further term which arises from the $Q$-variation of the $P(x;\lambda,\rho)$
factor in the primed commutator. This yields
\beq
-\frac{1}{2}\chi^\dagger_{\mu\nu}[W_{\mu\lambda\rho},W_{\nu\lambda\rho}]Q\left(P(x+\mu+\nu;
\lambda\rho)\right)\eeq
This operator will vanish in the naive continuum limit as expected.

Consider now the twisted lattice fermion action $S_F+S_Y$. In the continuum this
can be recast in terms of a single \KD field $\Psi$. 
\beq
S_{F+Y}=\Psi^\dagger M(U,\phi,V)\Psi\eeq
On the lattice, it is
convenient, although {\it not essential} to 
to rewrite it in terms of a single {\it complex} \KD field $\Psi$. 
However the following
Yukawa coupling is problematic in this respect as it involves a
$\Psi^\dagger\Psi^\dagger$ coupling
\beq
\chi^\dagger_{\mu\nu}(x)[\psi^\dagger_\lambda(x),W_{\lambda\mu\nu}(x)]\eeq
Such a term could be accommodated in the lattice theory by doubling the
number of fermion fields (and taking an appropriate further square
root of the fermion determinant when calculating the
effective bosonic action). Such an approach has the merit
of preserving the $Q$-symmetry exactly. Alternatively,
one can get around this problem
by replacing this term with the another similar one
\beq
\chi^\dagger_{\mu\nu}
[U^\dagger_\lambda(x)\psi_\lambda(x)U^\dagger_\lambda(x),W_{\lambda\mu\nu}(x)]\eeq
This modification of the lattice theory corresponds to a soft breaking
of the twisted SUSY and as such should not lead to dangerous SUSY violating
corrections. The additional soft breaking terms
would then vanish in the continuum limit. Notice that a $Q$-violating gluino
mass
term is likely to be necessary to control any numerical algorithm
used to simulate this theory and such a term will look rather
like the modified Yukawa we discuss here.
However, further work will be
needed to completely clarify this issue, in particular, how the
presence of such a soft breaking term effects the other twisted
supersymmetries.

Finally, using this latter
prescription leads, after integration over the complex twisted fermions, 
to a factor of ${\rm det}M(U,\phi,V)$ in the effective action for
the bosonic fields.

\subsection{Continuum Limit}
We have formulated the lattice theory in terms of a set of complex fields
since in this way we can preserve both gauge invariance and the twisted
supersymmetry. However to target ${\cal N}=4$ super Yang-Mills we
would like to restrict the path integrals defining correlation
functions to the real line (actually, in the case of the scalars
we need to impose the condition $\phib=-\phi^\dagger$). 
The first point to note is that the on-shell bosonic action $S_B$ is still
gauge invariant if the imaginary parts of $W_{\mu\nu\lambda}$, $U_\mu$
are set to zero. Furthermore, as in the continuum theory, we will
assume that the correct weight for the lattice twisted fermions after projection
to the real line is just the Pfaffian
of the full \KD operator (including the Yukawas). The latter 
can be replaced by ${\rm det}^{\frac{1}{2}}M(U,\phi,V)$ which is
then also gauge invariant. Finally, it should be clear that the lattice
action we have proposed reduces in the naive continuum limit to the usual super Yang-Mills
form along the real line. So the remaining question is whether the
Ward identities corresponding to the twisted supersymmetry $Q$ continue
to hold along this contour. We have no proof of this statement but can make
the following plausibility argument.

The lattice action is $Q$-exact which implies that any $Q$-invariant
observable can be computed in the limit $\beta\to\infty$. This limit
corresponds to the naive continuum limit in which the gauge links
are expanded to first order in the complex
gauge field $C_\mu=A_\mu+iB_\mu$ and lattice
expressions can be identified with their continuum counterparts. 
Consider the (complexified) Yang-Mills field strength that arises in
such a limit
\beq \beta\left[\left(F_{\mu\nu}(A)-[B_\mu,B_\nu]\right)^2+
          \left(D_{\left[\mu\right.} B_{\left.\nu\right]}\right)^2\right]\eeq
As $\beta\to\infty$ it is clear that the path integral is saturated by
fields with $B_\mu=0$ and $F_{\mu\nu}=0$. Thus it should be
possible to compute any $Q$-invariant
observable on the subspace corresponding to $B_\mu=0$. The Ward identities
are generated by picking a trivial $Q$-invariant observable, namely
$QO(\Psi,\Phi)$. We can thus conjecture that these Ward identities will
hold {\it on the real line} at finite lattice spacing. This conjecture should
be checked by perturbative calculations and numerical simulation. 

\section{Conclusions}
In this paper we propose a discretization of the ${\cal N}=4$
twisted Yang-Mills
action in four dimensions which is a generalization of the procedure used earlier to
construct a lattice theory of ${\cal N}=2$ super Yang-Mills theory in
two dimensions \cite{2dsuper}. 
The approach emphasizes the geometrical character of the twisted theory -- the twist
of ${\cal N}=4$ that we consider contains only integer spin fields and the
fermion content is naturally embedded in a (real) \KD field with anticommuting
components. This
decomposition of the fermions
also implies the existence of a scalar supercharge $Q$ which is
nilpotent up to gauge transformations. The superpartners of
the fermions are now contained in another \KD field. We write down a simple
set of transformation rules for the fields under this supercharge.
A $Q$-exact action is derived and shown to reduce to a well-known
twisting of ${\cal N}=4$ super Yang-Mills after a suitable
change of variables. For
completeness we also show how the original spinor formulation
of the theory can be recovered from this twisted model.

This manifestly geometric starting point allows us
to discretize the theory without inducing spectrum doubling and
maintaining both gauge invariance and a single twisted supersymmetry.
The lattice theory naturally contains complex fields -- to access
the correct continuum limit requires that an
appropriate contour be chosen when evaluating the path integral.
We argue that both gauge invariance and the twisted supersymmetry
can be maintained if this contour is chosen such that the
imaginary parts of all component fields bar the scalars are taken
to vanish. The scalars $\phi^a(x)$ and $\phib^a(x)$ are taken
to be complex conjugates of each other. The resulting effective action
is real, positive semidefinite at least for small enough lattice
spacing.

There are several directions for further work. 
The most obvious is the need for both perturbative and numerical checks
on the twisted supersymmetric Ward identities
corresponding to $Q$. If the results of those are
positive it will be necessary to derive and examine the Ward
identities following from additional elements of the twisted
superalgebra. 
In general we would expect that these latter Ward
identities would be broken at finite lattice
spacing. The hope is that the presence of a single
exact supersymmetry will, however, largely prohibit the theory from
developing relevant operators breaking these additional
supersymmetries. In this context it will be crucial
to pursue further studies, both analytical and
numerical, directed at allowing us to understand 
what, if any, additional
fine tuning is needed to reach the full ${\cal N}=4$ theory in the
continuum limit.

\acknowledgments
This work
was supported in part by DOE grant DE-FG02-85ER40237. The author
would like to acknowledge useful discussions with David Adams and Sofiane Ghadab.


\begin{thebibliography}{99}
\bibitem{2dsuper} S. Catterall, JHEP 0411 (2004) 006.
\bibitem{old} M Golterman and D. Petcher, Nucl. Phys. B319 (1989) 307.\\
S. Elitzur and A. Schwimmer, Nucl. Phys. B226 (1983) 109.\\
N. Sakai and M. Sakamoto, Nucl. Phys. B229 (1983) 173.\\
J. Bartels and J. Bronzan, Phys. Rev. D28 (1983) 818\\
T. Banks and P. Windey, Nucl. Phys. B198 (1983) 68.\\
H. Aratyn and A.H. Zimerman, J. Phys. A 18 (1985) L487.
(1985) 225\\
Y. Kikukawa and Y. Nakayama, Phys. Rev. D66 (2002) 094508.\\
K. Fujikawa, Phys. Rev. D66 (2002) 074510.\\
J. Nishimura, Phys. Lett. B406 (1997) 215.\\
S. Catterall and S. Karamov, Phys. Rev. D 68 (2003) 014503.\\
W. Bietenholtz, Mod. Phys. Lett. A14 (1999) 51.\\
M. Beccaria, M. Campostrini, G.De Angelis and A. Feo, Phys. Rev. D70 (2004)\\
M. Beccaria, M. Campostrini, A. Feo, Phys. Rev. D69 (2004) 095010.
035011.
\bibitem{feo_rev} A. Feo, Nucl. Phys. 119 (Proc. Suppl.) (2003) 198.
\bibitem{kap_rev} D. B. Kaplan, Nucl. Phys. B (Proc. Suppl.) 129-130 (2004) 109.
\bibitem{kap}
D.B. Kaplan, E. Katz and M. Unsal, JHEP 0305 (2003) 037, A.G. Cohen, D.B. Kaplan, E. Katz, M. Unsal, 
JHEP 0308 (2003) 024, A.G. Cohen, D.B. Kaplan, E. Katz, M. Unsal, JHEP 0312
(2003) 031\\
\bibitem{others}
J. Nishimura, S. Rey and F. Sugino, JHEP 0302 (2003) 032\\
J. Giedt, E. Poppitz and M. Rozali, JHEP 0303 (2003) 035\\
J. Giedt, Nucl. Phys. B668 (2003) 138\\
J. Giedt, Nucl.Phys. B674 (2003) 259.
\bibitem{feowz} M. Bonini and A. Feo, JHEP 0409 (2004) 011.
\bibitem{moto} $N=1$ super Yang-Mills on a (3+1) dimensional 
transverse lattice, M. Harada, S. Pinsky, hep-lat/0411024, M. Harada, S. Pinsky,
Phys. Rev. D70 (2004) 087701, M. Harada, S. Pinsky, Phys. Lett. B567 (2003) 277.
\bibitem{top} S. Catterall, JHEP 0305 (2003) 038.
\bibitem{qm} S. Catterall and E. Gregory, Phys. Lett. B487 (2000) 349.
\bibitem{wz2} S. Catterall and S. Karamov, Phys. Rev. D65 (2002) 094501.
\bibitem{sigma} S. Catterall and S. Ghadab, JHEP 0405 (2004) 044.
\bibitem{sug} F. Sugino, JHEP 0401 (2004) 015, F. Sugino, JHEP 0403 (2004) 067,
F. Sugino, JHEP 0501 (2005) 016.
\bibitem{noboru} A. D'Adda, I. Kamamori, N. Kawamoto, K. Nagata Nucl.
Phys. B707 (2005) 100. \\
N. Kawamoto , T. Tsukioka, Phys. Rev. D61 (105009) 2000.\\
J. Kato, N. Kawamoto and Y. Uchida, Int. J. Mod. Phys. A19 (2004) 2149.\\
J. Kato, N. Kawamoto and A. Miyake, hep-th/0502119
\bibitem{schwim} S. Elitzur, E. Rabinovici, A. Schwimmer, Phys. Lett. B119 (1982) 165.
\bibitem{witten} E. Witten, Comm. Math. Phys. 117 (1988) 353.
\bibitem{conf} S. Catterall, Nucl. Phys. B Proceedings Suppl. 140 (2005) 751. 
\bibitem{banks} T. Banks, Y. Dothan and D. Horn, Phys. Lett. B117 (1982) 413.
\bibitem{rabin} J. Rabin, Nucl. Phys. B201 (1982) 315.
\bibitem{betch} P. Becher, Phys. Lett. B104 (1981) 221.
\bibitem{joos} P. Becher and H. Joos, Z. Phys. C 15 (1982) 343.
\bibitem{marcus} N. Marcus, Nucl. Phys. B452 (1995) 331.
\bibitem{lambastida} J. Labastida and C. Lozano, Nucl. Phys. B502 (1997) 741.
\bibitem{blau} M. Blau and G. Thompson Nucl. Phys. B492 (1997) 545.
\bibitem{adjoint} H. Aratyn, M. Goto and A.H. Zimerman, Nuovo Cimento A84,
(1984) 255
\bibitem{geo} H. Aratyn and A.H. Zimerman, Phys. Rev. D33 (1986) 2999.
\end{thebibliography}
\end{document}